# Cross-sectional topological anomaly scores and intraday return predictability in the S&P 500: A BallMapper, decoder-conditional VAE, and Function-on-Function regression approach

Krzysztof Ozimek[*]

## Abstract

Anomaly detection methods in financial time series score statistically unusual observations in observable data, not topologically misexpected persistent deviations in the latent structure of co-movement. This study constructs a stock-level topological anomaly score jointly conditioned on market-level topological structure and cross-sectional peer context, and tests whether its history carries predictive content for return curves. Intraday data for ten liquid S&P 500 constituents (April 2025–March 2026) are embedded via Takens delay embedding, graphed by BallMapper, and scored by three decoder-conditional variational autoencoder variants. Predictive content is assessed by penalised function-on-function regression and confirmed across all assets, intraday bar frequencies, and scoring variants, revealing a consistent temporal fingerprint — gradual accumulation of return impact, a frequent early reversal of its direction, and broadly distributed predictive content weighted toward recent anomaly history. When the reversal occurs depends on market regime; how evenly the anomaly history contributes to prediction depends on bar frequency.

[*]e-mail: contact@drkrzysztofozimek.com

## 1. Introduction

The efficient market hypothesis (EMH), as formalised by Fama (1970), holds that security prices fully and instantaneously incorporate all available information, leaving no systematic opportunity for risk-adjusted excess returns. Yet decades of empirical research have documented persistent deviations from this benchmark. Calendar anomalies — the January effect, the day-of-the-week effect, and the turn-of-the-month effect — cross-sectional regularities including the size, value, and momentum premia (Fama, French 1993; Jegadeesh, Titman 1993), and return dynamics at both short and long horizons, including the overreaction evidence of De Bondt and Thaler (1985) and the mean-reversion findings of Poterba and Summers (1988), have collectively challenged the unconditional efficiency paradigm and sustained academic interest in detecting and quantifying anomalous market behaviour.

The established financial-econometric methodology for anomaly identification rests on two pillars: multifactor models and event studies. In the factor-model framework, an anomaly manifests as a statistically significant intercept — an *alpha* — in a regression of excess returns on a benchmark factor set; the progression from the single-factor CAPM through the Fama–French three- and five-factor models (Fama, French 1993; Fama, French 2015) to contemporaneous machine-learning factor zoos reflects an ongoing search for anomalies that survive risk adjustment. Event studies measure price responses to specific informational events by computing cumulative abnormal returns relative to a market model. Both approaches operate at daily or event-window resolution and are designed for *pre-specified* anomaly types — calendar-based, factor-based, or event-driven. Neither identifies anomalies that arise from the latent geometric structure of intraday cross-sectional co-movement: transient configurations of the joint return distribution that may carry forward-looking information yet defy characterisation through parametric factor loadings or discrete event catalogues.

A complementary strand of research on algorithmic event and anomaly detection in time series offers a richer, more systematic toolkit. Mehrotra, Mohan and Huang (2017) establish the foundational taxonomy: distance- and density-based methods (k-nearest-neighbour detectors, local outlier factor, INFLO, LOCI), clustering-based methods, model-based methods (regression residuals, hidden Markov models, ARIMA filters, feedforward neural networks), and ensemble approaches. Ogasawara, Salles, Porto and Pacitti (2025), in a comprehensive tutorial survey covering the full event-detection spectrum — anomalies, change points, and motifs — also cover deep-learning detectors (convolutional DeepAnT, DeepNAP, LSTM-based autoencoders, Telemanom), spectral and information-theoretic methods, change-point tests (CUSUM, Chow, PELT), and graph-based spatiotemporal neural networks for multivariate series; they further encompass large-language-model-based forecasters as an emerging frontier. Together, these works represent the state of the art across the full breadth of modern ML and AI approaches to anomaly detection. Yet all of them — from classical LOF through transformer architectures — share one conceptual foundation: an anomaly is a *temporal or statistical outlier* in an individual asset's own series or, at most, in pairwise inter-asset correlations. Ogasawara et al. (2025) do introduce the important distinction between *contextual anomalies* (atypical only within a temporal neighbourhood) and *out-of-context anomalies*, which is conceptually the closest notion to our approach; however, their context is purely temporal — defined by surrounding bars in a single series — rather than structural, defined by an asset's geometric position within a cross-sectional co-movement graph. The shape of the space in which all assets collectively move at each intraday snapshot lies outside the scope of any of these frameworks.

Topological data analysis (TDA) offers a qualitatively different perspective: rather than fitting a distribution and measuring deviations from it, TDA studies the intrinsic shape of data — how points cluster, connect, and separate — without imposing parametric assumptions (Carlsson 2009; Edelsbrunner, Harer 2010). Applied to financial markets, persistent-homology methods have detected early warning signals of crises (Gidea, Katz 2018) and enhanced financial time series forecasting accuracy (de Jesus Jr. et al. 2024), while Shiraj et al. (2024) applied Mapper with DBSCAN clustering to detect anomalies in equity prices. Notwithstanding this progress, three gaps persist in the literature. First, existing TDA-in-finance studies characterise the *aggregate* topology of a market-level point cloud but do not localise individual assets within the graph or score their structural position in a peer-context-aware, unsupervised manner; no stock-level topological anomaly score has been proposed. Second, while the regression-residual and deep-learning reconstruction frameworks surveyed by both Mehrotra et al. (2017) and Ogasawara et al. (2025) score anomalousness relative to a model fitted to a single asset's own history, neither conditions the anomaly score on the *cross-sectional structural configuration* of all peers simultaneously — the conditioning that is essential when the anomaly of interest is geometric isolation from the co-movement graph rather than marginal tail behaviour of one series. Third, the predictive content of TDA-derived anomaly scores for subsequent intraday return *curves* — not merely point-in-time returns — has not been assessed using the functional data analytical tools naturally suited to quantifying the full time-profile of a predictive relationship (Ramsay, Silverman 2005; Ivanescu et al. 2015).

The remainder of the paper is organised as follows. Section 2. states the research goals and formally specifies the three hypotheses. Section 3. describes the data and the three-stage methodology: the topological pipeline (Section 3.2.), the decoder-conditional VAE (Section 3.3.), and the function-on-function regression framework (Section 3.4.). Section 4. presents and interprets the results. Section 5. acknowledges the principal limitations. Section 6. concludes with practical implications and directions for future research.

## 2. Goals and Hypotheses

This paper addresses all three gaps. Its threefold objective is: **(i)** to develop a topological pipeline — grounded in Takens delay embedding (Takens 1981; Sauer, Yorke, Casdagli 1991) and the BallMapper algorithm (Dłotko 2019) — that produces market-level and stock-level structural descriptors from intraday S&P 500 return point clouds of the ten sample stocks; **(ii)** to design and compare three variants of a decoder-conditional variational autoencoder (VAE) that assigns, in an unsupervised fashion, a peer-context-aware structural anomaly score to each stock at each intraday snapshot (Kingma, Welling 2014; An, Cho 2015); and **(iii)** to assess whether the temporal history of these scores carries statistically and economically significant information about subsequent cumulative returns, estimated via penalised function-on-function regression (Ivanescu et al. 2015).

An anomaly, as used in this paper, is any mismatch between the intraday return pattern expected for a stock — given the market's current return structure as seen topologically and the stock's peer group — and what is actually observed, regardless of the magnitude or direction of the departure. The history of such mismatches serves as the predictor of subsequent cumulative returns. The three hypotheses below concern three qualitative aspects of how that predictive relationship is shaped: the temporal persistence of the return impact across the forecast horizon, the distribution of predictive content across the anomaly history window, and the directional shape of the return response. Each hypothesis is aligned with a well-established stylized fact documented across multiple timescales; since the topological anomaly setting is novel, however, they are treated as informed rather than strictly confirmatory predictions. The corresponding scalar measures are formally defined in Section 3.4.

**Hypothesis concerning response side: how the return impact dissipates**. **$H_1$**: *The cumulative return impact of an anomaly episode accumulates gradually and persistently across the forecast horizon.* This is consistent with the limits-to-arbitrage stylized fact (Shleifer, Vishny 1997): capital constraints and risk aversion of institutional arbitrageurs slow the full absorption of structural mispricings, producing gradual rather than instantaneous return correction.

**Hypothesis concerning predictor side: which part of the anomaly history matters**. **$H_2$**: *The predictive content of the anomaly history is distributed broadly across the full history window, with a statistically significant lean toward more recent observations.* This reflects the information-decay principle: the predictive relevance of a past signal diminishes as market conditions evolve, so recent anomaly readings carry disproportionately higher weight — a regularity implicit in time-varying conditional variance models across all frequencies (Engle 1982).

**Hypothesis concerning shape of response: whether the return effect reverses direction**. **$H_3$**: *The cumulative return impact of an anomaly episode typically reverses direction before the end of the forecast horizon.* This is grounded in two complementary stylized facts: the overreaction-and-correction regularity documented at medium-to-long horizons (De Bondt, Thaler 1985), and the liquidity-provision mechanism by which market makers absorb order-flow imbalances and allow prices to revert as inventory is unwound (Grossman, Miller 1988).

The paper makes three contributions. It introduces the first stock-level topological anomaly score constructed from a dynamic BallMapper cross-sectional graph, advancing TDA-in-finance beyond the aggregate market-topology analyses reviewed in the TDA literature. It proposes a novel decoder-conditional VAE architecture in which the decoder is jointly conditioned on a market-level topological descriptor and three peer-context encodings — raw, aggregate, and attention-weighted — allowing systematic comparison of context specification strategies, and directly addressing the absence of market-topology and cross-sectional peer conditioning in the algorithmic anomaly-detection frameworks of Mehrotra et al. (2017) and Ogasawara et al. (2025). It establishes a functional regression link between the history of structural anomaly scores and forward return curves, quantifying the full time-profile of the return impact through five interpretable scalar measures — the front-load ratio, the median effect horizon, the predictor centre of mass, the uniformity ratio, and the zero-crossing horizon — none of which is available in standard event-detection or return-anomaly frameworks.

# 3. Data and Methodology

The three-stage analytical framework is illustrated schematically in Figure 1 (Appendix A.0.). The following subsections describe each stage in detail.

## 3.1. Data

The study uses intraday OHLCV (Open/High/Low/Close/Volume) data for ten highly liquid S&P 500 stocks over the twelve-month period 2025-04-01 – 2026-03-31. The ten constituents — NVDA, AAPL, TSLA, AMZN, META, MSFT, GOOG, GOOGL, AVGO, and PLTR — were selected from the full index by ranking all constituents on the Amihud (2002) illiquidity ratio computed from daily price and volume data sourced from Yahoo Finance; the ten stocks with the lowest (most liquid) ratio were retained. The full selection procedure and liquidity table are given in Appendix A.1. (Table 1). Although GOOG (Class C) and GOOGL (Class A) share the same underlying issuer, at intraday frequency price formation — bid-ask spreads, order-flow composition, and tick-by-tick dynamics — is share-class-specific rather than issuer-specific, so their delay-embedding vectors $\boldsymbol{\varphi}_{i,k}$ (Section 3.2.) will generically occupy different positions in $\mathbb{R}^3$ at each snapshot. Their joint inclusion is a direct and unmodified consequence of applying the Amihud filter mechanically to all index constituents; no discretionary adjustment was made.

The sample window spans twelve calendar months, determined by the availability of EODHD (eodhd.com) intraday data and the requirement of the factorial design for four complete calendar quarters. The resulting period encompasses a sequence of temporally distinct yet economically connected geopolitical and macro-financial stress regimes affecting U.S. equity markets — an observed property of the window rather than a selection criterion. These include the broad U.S. tariff announcement of 2 April 2025 and the subsequent trade-policy escalation, continued geopolitical instability associated with the Russia–Ukraine conflict, military strikes on Iranian nuclear infrastructure in June 2025, and the renewed escalation of U.S.–Iran–Israel tensions during Q1 2026. Collectively, these events generated repeated episodes of volatility clustering, sectoral repricing, liquidity shifts, and abrupt changes in intraday market dynamics. The resulting horizon therefore provides a structurally heterogeneous environment suitable for evaluating the robustness of the anomaly-detection pipeline under multiple forms of market stress rather than under a single isolated event. Importantly, the selected period contains both acute shock episodes and intermediate stabilisation phases, allowing the analysis to capture transitions between elevated and relatively normal market conditions. This characteristic is particularly relevant for studying anomaly persistence, temporal propagation, and cross-sectional dependence structures in high-frequency financial data.

Five-minute OHLCV bars for the selected stocks were obtained from EODHD. The NASDAQ regular session (09:30–16:00 ET, 78 bars per day, with bar opens from 09:30 to 15:55) forms the reference grid; market holidays are excluded. Missing bars were imputed by a three-tier scheme: LOCF (Last Observation Carried Forward) for single-bar gaps, a Kalman smoother on close log-returns for gaps of 2–5 bars, and cross-sectional OLS regression on the nine remaining stocks for longer gaps; volume gaps were handled by LOCF throughout. After imputation, OHLC geometry constraints $L_t \leq \min(O_t, C_t) \leq \max(O_t, C_t) \leq H_t$ were enforced on every bar.

The working variable throughout the analysis is the intrabar log return

$$r_{i,t} = lnC_{i,t} - lnO_{i,t}, \tag{1}$$

where $C_{i,t}$ and $O_{i,t}$ are the close and open prices of stock $i$ at bar $t$. The five-minute bars (5M) are aggregated to 15M, 30M, and 60M timeframes by standard OHLCV rules; all four frequencies are partitioned into four calendar quarters (Q2 2025 – Q1 2026) that serve as replication units in Section 4. Distributional diagnostics confirm non-normality, stationarity, and volatility clustering across all assets and frequencies, with Geweke–Porter–Hudak (GPH) test rejections that are episodic and frequency-specific rather than systematic, supporting treatment of all series as effectively short-memory.

## 3.2. Dynamic Topological Description

### Takens Delay Embedding

Each intraday snapshot $k$ is indexed by its *anchor* $a_k$ — the bar index of the most recent return entering the delay vector at that snapshot. With embedding dimension $m = 3$ and time delay $\tau = 1$ the anchor advances as $a_k = k + 2$: snapshot $k = 1$ anchors at bar 3, snapshot $k = K$ anchors at bar $T$, giving $K = T - 2$ snapshots in total. The delay-embedding vector for stock $i$ at snapshot $k$ is

$$\boldsymbol{\varphi}_{i,k} = (r_{i,a_k-2},\ r_{i,a_k-1},\ r_{i,a_k})^{\top} \in \mathbb{R}^3, \tag{2}$$

ordered from oldest (leftmost) to most recent (rightmost). Stacking all $N = 10$ vectors at snapshot $k$ yields the cross-sectional cloud matrix $\boldsymbol{C}_k \in \mathbb{R}^{N\times 3}$, which serves as input to BallMapper.

### BallMapper and Snapshot Graph Construction

At each snapshot $k$, BallMapper covers $\boldsymbol{C}_k$ with overlapping balls of radius $\hat{\varepsilon}$, selected via a Median Absolute Deviation (MAD)-based Gaussian approximation with nominal coverage probability $\alpha = 0.68$. Specifically, $\hat{\varepsilon} = \Phi^{-1}\left(\frac{1+\alpha}{2}\right) \cdot med(d)/\Phi^{-1}(0.75)$, where $\Phi^{-1}$ is the standard-normal quantile function, $med(d)$ is the median of the pairwise distances pooled from a representative sample of the clouds, an $\Phi^{-1}(0.75) \approx 0.6745$ is the MAD-to-$\hat{\sigma}$ consistency factor under normality (for $X \sim \mathcal{N}(0, \sigma^2)$, median $(|X|) = \sigma\Phi^{-1}(0.75)$, so dividing by 0.6745 recovers $\hat{\sigma}$); with $\alpha = 0.68$ (one standard deviation rule), $\Phi^{-1}(0.84) \approx 0.9945$, giving $\hat{\varepsilon} \approx 0.9945 \times med(d)/0.6745$. It is equivalent to setting the radius to approximately one estimated standard deviation of the cloud's pairwise distance distribution. Two balls are connected by an edge whenever their stock-membership sets overlap, yielding the snapshot graph $G_k = (V_k, E_k)$ with one node per ball; individual stocks may belong to more than one ball, so the graph simultaneously encodes cluster membership and cross-cluster boundary positions.

**Topological Graph Descriptors**

For each snapshot $k$, two complementary sets of descriptors are extracted from the snapshot graph $G_k$. The first set characterises the topology of the graph as a whole and serves as a market-level summary of the S&P 500 cross-sectional structure at time $k$. The second set locates each individual stock within the graph and provides a stock-level measure of structural position. Together, these two sets yield time series of topological features over the full sample of $K$ snapshots.

The market-level descriptor vector at snapshot $k$ is the four-dimensional row vector

$$\boldsymbol{\Gamma}(k) = (B(k), Z(k), C(k), \bar{d}(k)), \quad (3)$$

whose components are formally defined in Table 2 (Appendix A.2.). Stacking $\boldsymbol{\Gamma}(1), \dots, \boldsymbol{\Gamma}(K)$ by rows forms the $K \times 4$ matrix $\boldsymbol{\Gamma}$.

The stock-level descriptor vector for asset $i$ at snapshot $k$ is built from the set of balls in $G_k$ that contain asset $i$; because balls may overlap, a stock can simultaneously belong to more than one ball. The descriptor vector is

$$\boldsymbol{\Phi}_i(k) = (M_i(k),\ I_i(k),\ D_i^{max}(k),\ \bar{D}_i(k)), \quad (4)$$

whose components are defined in Table 3 (Appendix A.2.). Stacking $\boldsymbol{\Phi}_i(1), \dots, \boldsymbol{\Phi}_i(K)$ by rows forms the $K \times 4$ matrix $\boldsymbol{\Phi}_i$ for each $i = 1, \dots, N$.

All eight descriptors — four market-level ($B$, $Z$, $C$, $\bar{d}$) and four stock-level ($M_i$, $I_i$, $D_i^{\max}$, $\bar{D}_i$) — are non-normal (JB rejects universally), stationary (ADF rejects at conventional levels across all quarters and timeframes), and exhibit volatility clustering (ARCH rejects at 1% for all market-level descriptors and in the large majority of stock-level cases). GPH rejections are sparse and unsystematic for both sets, confirming that the descriptor sequences are effectively short-memory — the same characterisation established for the underlying log returns in Section 3.1.

## 3.3. Anomaly Detection via Decoder-Conditional VAE

The topological descriptor sequences produced in Section 3.2. — the $K \times 4$ market-level matrix $\boldsymbol{\Gamma}$ and the collection $\{\boldsymbol{\Phi}_i \in \mathbb{R}^{K \times 4}\}_{i=1}^{N}$ of asset-level matrices — provide a continuous record of each of the ten sampled S&P 500 constituent's structural position within the BallMapper graph across the $K$ intraday snapshots of the sample period. The objective of this section is to convert these sequences into a scalar anomaly score $A_i(t)$ that, for each asset $i$ and each snapshot $t$, quantifies how surprising asset $i$'s structural position is relative to the prevailing market topology and the contemporaneous cross-section of peer positions.

**Data Preparation and Standardisation**

$\boldsymbol{\Gamma}$ and each $\boldsymbol{\Phi}_i$ are standardised independently to zero mean and unit variance using statistics computed from the full $K$-snapshot sample. A shared, asset-specific, context-free encoder $q_{\boldsymbol{\phi}}(\boldsymbol{z} \mid \boldsymbol{\Phi}_i(t)) = \mathcal{N}(\boldsymbol{\mu}_{\boldsymbol{\phi}}(\boldsymbol{\Phi}_i(t)), diag(\boldsymbol{\sigma}^2_{\boldsymbol{\phi}}(\boldsymbol{\Phi}_i(t))))$, where $\boldsymbol{\phi}$ collects the learnable parameters of the encoder network, maps each stock's standardised descriptor to a Gaussian distribution over a continuous latent variable $\boldsymbol{z}$; a context-conditional $\beta$-VAE decoder $\widehat{\boldsymbol{\Phi}}_i(t) = \psi_{\boldsymbol{\Theta}}(\boldsymbol{z}, \boldsymbol{c}_i(t), \boldsymbol{\Gamma}(t))$, where $\boldsymbol{\Theta}$ collects the learnable parameters of the decoder network, then reconstructs it jointly from the latent code, the peer context $\boldsymbol{c}_i(t)$, and the market-level descriptor $\boldsymbol{\Gamma}(t)$. The three variants described below differ only in how $\boldsymbol{c}_i(t)$ is formed; all are trained by maximising the $\beta$-VAE evidence lower bound (ELBO; Higgins et al. 2017) $\mathcal{L} = \mathbb{E}_{q_{\boldsymbol{\phi}}(\boldsymbol{z}|\boldsymbol{\Phi}_i(t))}[log\ p_{\boldsymbol{\Theta}}(\boldsymbol{\Phi}_i(t) \mid \boldsymbol{z}, \boldsymbol{c}_i(t))] - \beta D_{KL}\big(q_{\boldsymbol{\phi}}(\boldsymbol{z} \mid \boldsymbol{\Phi}_i(t)) \| \mathcal{N}(\mathbf{0}, \boldsymbol{I})\big)$, where $\beta > 0$ is the KL penalty weight and $D_{KL}$ denotes the Kullback–Leibler (KL) divergence.

**Full Raw Peer Context (VAE-I)**

VAE-I constructs the peer context as the raw stack of all co-listed stock descriptors: $\boldsymbol{c}_i^{(\mathbf{1})}(t) = \left[\boldsymbol{\Gamma}(t), \{\boldsymbol{\Phi}_j(t)\}_{j \neq i}\right] \in \mathbb{R}^{4N}$, stacking the descriptor vectors of all $N-1$ peers without aggregation. This order-sensitive encoding is unshared - each asset receives a distinct peer matrix - and imposes no structure on bilateral peer relationships, capturing the full idiosyncratic cross-sectional configuration.

**Aggregate Peer Context (VAE-II)**

VAE-II summarises peer information via the cross-sectional mean and standard deviation: $\boldsymbol{c}_i^{(2)}(t) = \left[\boldsymbol{\Gamma}(t), \frac{1}{N-1}\sum_{j \neq i} \boldsymbol{\Phi}_j(t), \boldsymbol{s}_i(t)\right] \in \mathbb{R}^{12}$, where $\boldsymbol{s}_i(t) \in \mathbb{R}^4$ is the element-wise cross-sectional standard deviation over the $N-1$ peers. This compact, permutation-invariant encoding captures both the average level and the dispersion of peer-group structure while eliminating the ordering ambiguity of VAE-I, and applies the same aggregation rules symmetrically across all assets.

**Attention-Based Peer Context (VAE-III)**

VAE-III computes a content-based weighted peer summary: $\boldsymbol{c}_i^{(3)}(t) = \left[\boldsymbol{\Gamma}(t), \sum_{j \neq i} \alpha_{ij}(t)\boldsymbol{\Phi}_j(t)\right] \in \mathbb{R}^8$, where attention weights $\alpha_{ij}(t) = softmax_j\big(\boldsymbol{w}^\top \boldsymbol{\Phi}_j(t)\big)$ are derived from a learnable linear scoring of each peer's descriptor alone, with $\mathrm{w} \in \mathbb{R}^4$ a shared weight vector, allowing the effective peer context to adapt dynamically to the current cross-sectional configuration in a permutation-equivariant fashion.

**Anomaly Score**

After training, the anomaly score for asset $i$ at snapshot $t$ is computed deterministically — the stochastic latent sample is replaced by its mean, $\boldsymbol{z} = \boldsymbol{\mu}_{\phi}(\boldsymbol{\Phi}_i(t))$ — and combines the Euclidean reconstruction residual with the KL divergence of the encoder posterior:

$$A_i(t) = \| \boldsymbol{\Phi}_i(t) - \widehat{\boldsymbol{\Phi}}_i(t) \|_2 + \beta \cdot D_{KL}\big(q_\phi(\boldsymbol{z} \mid \boldsymbol{\Phi}_i(t)) \ \| \ \mathcal{N}(\mathbf{0}, \boldsymbol{I})\big), \tag{5}$$

where both terms are evaluated in the standardised feature space. The first term measures *structural surprise*: how far the reconstructed descriptor $\widehat{\boldsymbol{\Phi}}_i(t)$ — produced by the decoder given the latent code and the peer context — deviates from the observed descriptor $\boldsymbol{\Phi}_i(t)$ in the four-dimensional topological feature space. The second term measures *latent irregularity*: how far the encoder's inferred posterior for asset $i$ deviates from the standard normal prior, capturing whether the stock's topological position is intrinsically atypical independently of the market context. Stacking scores over all assets and snapshots yields the score matrix $\boldsymbol{A} \in \mathbb{R}^{K \times N}$.

## 3.4. Penalised Function-on-Function Regression

### The Anchored Function-on-Function Model

In the present application, the replications are BallMapper anchor snapshots $a_k$, $k = 1, \ldots, K$. For replication $k$ the predictor function is the anomaly score history of asset $i$ over the $W$ observations immediately preceding the anchor, with the predictor domain standardised to $[0, W]$ by setting $w = 0$ at the oldest observation and $w = W$ at the anchor snapshot itself. The response function is the cumulative log-return of asset $i$ from the anchor to each subsequent horizon: $r_i(a_k, a_k + h) = \log P_i(a_k + h) - \log P_i(a_k)$, $k = 1, \ldots, K$; $h = 1, \ldots, H$. The anchored function-on-function model is

$$r_i(a_k,\, a_k + h) \;=\; \beta_0(h) \;+\; \int_0^W \beta\,(w, h)\, A_i(w)\, dw \;+\; \varepsilon_i(a_k, h), \tag{6}$$

where $A_i(w)$ is the anomaly score from equation (5) evaluated at relative past time $w \in [0, W]$ within the trailing window of replication $k$. The intercept function $\beta_0(h)$ captures the baseline cumulative return profile averaged across replications and anomaly histories. The bivariate coefficient surface $\beta(w, h)$, defined on $[0, W] \times [1, H]$, is the central inferential object: its value encodes how the anomaly level registered at relative past time $w$ within the history window influences the cumulative return at horizon $h$ after the anchor. The residuals $\varepsilon_i(a_k, h)$ are assumed independent across replications $k$ but may be correlated across horizons $h$ within a replication, consistently with the within-subject residual correlation structure discussed in Crainiceanu et al. (2024). A complete notation table is provided in Appendix A.4. (Table 5). The window width $W = 5$ trading days and forecast horizon $H = 2$ trading days are fixed on economic grounds; their derivation and translation to bar counts at each intraday frequency are detailed in Appendix A.3. (Table 4). Throughout this section $W$ and $H$ denote these integer bar counts (e.g., $W = 390$ for 5-minute bars); the $W$ discrete predictor observations at grid points $\{0,1, \ldots, W\}$ and the $H$ response observations at $\{1, \ldots, H\}$ are treated as samples from continuous functions on $[0, W]$ and $[1, H]$ respectively — the standard discrete-to-continuous passage in penalised FDA (Ramsay, Silverman 2005; Ivanescu et al. 2015).

The coefficient surface $\beta(w, h)$ is estimated by penalised function-on-function regression, which fits a tensor-product penalised spline surface to the $K$ functional data pairs. The penalty controls the roughness of $\beta(w, h)$ in both the $w$ and $h$ directions simultaneously, so that the estimated surface is smooth in economic time rather than artificially jagged at the discrete observation grid. This estimator, detailed in Ivanescu et al. (2015), produces pointwise confidence bands for $\beta(w,h)$ and permits a test of the global null hypothesis $H_0: \beta(w,h) = 0$ for all $(w,h) \in [0,W] \times [1,H]$.

**Quantifying the Anomaly Impact Pattern**

The estimated surface $\hat{\beta}(w,h)$ is a bivariate function that encodes simultaneously the temporal structure of how anomaly history enters return prediction and the horizon structure of how return impact evolves after the anchor. To reduce this surface to a set of economically interpretable statistics comparable across assets and the three anomaly score variants, five scalar measures are derived from two marginal summary functions.

The marginal effect at horizon $h$ integrates the coefficient surface over the full predictor domain, yielding the net effect of the complete anomaly history on the cumulative return at that horizon:

$$\bar{\beta}(h) = \int_0^W \beta(w,h)\, dw, \qquad h \in [1,H]. \tag{7}$$

The sign of $\bar{\beta}(h)$ indicates the direction of the impact at horizon $h$ and its magnitude indicates the aggregate strength. The total absolute effect attributable to the anomaly level at past time $w$ integrates the absolute surface over the full response domain:

$$\tilde{\beta}(w) = \int_1^H |\beta(w,h)|\, dh, \qquad w \in [0,W]. \tag{8}$$

$\tilde{\beta}(w)$ quantifies the total unsigned influence that the anomaly score at relative past time $w$ exerts on the return profile across all horizons. The cumulative share of the total return impact accumulated up to horizon $h_0$ is

$$\delta(h_0) = \frac{\int_1^{h_0} |\bar{\beta}(h)|\, dh}{\int_1^{H} |\bar{\beta}(h)|\, dh}, \qquad h_0 \in [1,H], \tag{9}$$

which is non-decreasing from $\delta(1) = 0$ to $\delta(H) = 1$ and has the interpretation of a cumulative distribution function of return-impact mass across the forecast horizon.

The five scalar measures derived from equations (7)–(9) each characterise one qualitative feature of the coefficient surface, arranged in two symmetric pairs plus one directional diagnostic: Measures 1 and 2 use $\delta(h_0)$ to describe the $h$-dimension — the concentration and location of the return impact across the forecast horizon; Measures 3 and 4 use $\tilde{\beta}(w)$ directly to describe the $w$-dimension — the centre of mass and spread of the predictor effect across the anomaly history window; Measure 5 uses $\bar{\beta}(h)$ to detect a directional reversal within the horizon.

*Measure 1:* **Concentration near small $h$ (front-load ratio).** The statistic $\delta(H/4)$ measures the proportion of the total return impact accumulated in the first quarter of the forecast horizon. Values substantially above $1/4$ indicate transient impact: the anomaly's effect on returns is front-loaded and dissipates quickly after the anchor. Values close to or below $1/4$ are consistent with persistent impact that continues to grow throughout the horizon window.

*Measure 2:* **Spread across large $h$ (median effect horizon).** The statistic $h_{0.50} = \min\{h: \delta(h) \geq 0.50\}$ is the horizon at which half the total absolute impact has accumulated — the half-life of the return effect. Small values of $h_{0.50}$ relative to $H/2$ confirm transient impact; values exceeding $H/2$ indicate that the impact is persistent, with more than half the total effect materialising in the second half of the forecast window. In conjunction with Measure 1, the pair $(\delta(H/4), h_{0.50})$ distinguishes four regimes: early-concentrated, late-concentrated, mid-concentrated, and uniformly distributed impact.

*Measure 3:* **Concentration near recent $w \approx W$ (centre of mass of predictor effect).** The statistic $\bar{w} = \int_0^W w\, \tilde{\beta}(w)\, dw / \int_0^W \tilde{\beta}\,(w)\, dw$ is the weighted average of relative past time with weights proportional to the total unsigned influence of the anomaly at that time. Values of $\bar{w}$ close to $W$ indicate that the most recent anomaly observations dominate predictive content, consistent with the recency-driven return patterns documented in the momentum literature (Jegadeesh, Titman 1993). Values near $W/2$ indicate that the anomaly history contributes approximately uniformly from distant to recent past.

*Measure 4:* **Spread across $w$ (uniformity ratio).** Let $\sigma_w^2 = \int_0^W (w - \bar{w})^2 \tilde{\beta}(w)\, dw / \int_0^W \tilde{\beta}\,(w)\, dw$ denote the dispersion of the predictor effect around its centre of mass. The uniformity ratio $U = \sigma_w^2/(W^2/12)$ normalises this dispersion by the variance of the uniform distribution on $[0, W]$. Values of $U$ near unity, when combined with $\bar{w} \approx W/2$, indicate that $\tilde{\beta}(w)$ is approximately uniformly distributed across the full history window, implying that every segment of the anomaly past contributes roughly equally to return prediction. Values of $U$ substantially below unity indicate concentration; the location of $\bar{w}$ then determines whether the concentration is at the recent or the distant end of the history window.

*Measure 5:* **Sign change across $h$ (zero-crossing horizon).** The statistic $h^* = \min\{h \in (1, H): \bar{\beta}(h) = 0 \text{ with a sign change}\}$ identifies the horizon at which the marginal return impact reverses direction. Its existence in the interior of $[1, H]$ is diagnostic of mean-reversion dynamics: the anomaly initially moves returns in one direction — typically depressing prices during the fragmentation phase — and subsequently reverses at $h = h^*$, consistent with the overreaction and price-correction mechanism documented by De Bondt and Thaler (1985) and the long-horizon mean-reversion evidence of Poterba and Summers (1988). When no sign change occurs within $[1, H]$, the impact is monotone and $h^*$ is not defined.

The five measures — $\delta(H/4)$, $h_{0.50}$, $\bar{w}$, $U$, and $h^*$ — are computed from the estimated surface $\hat{\beta}(w, h)$ after fitting equation (6) separately for each of the ten sampled S&P 500 constituents under each of the three anomaly score variants from Section 3.3. Their values across all assets and variants are reported in Section 4.

To test whether the five scalar measures vary systematically with calendar quarter ($Q$), bar timeframe ($F$), and anomaly-scoring method ($V$), each measure is analysed separately with a three-way crossed mixed model. In compact symbolic form the model is

$$y_i \;=\; \beta_0 + Q_i + F_i + V_i + QF_i + QV_i + FV_i + QFV_i + b_{A[i]} + \varepsilon_i, \tag{10}$$

where $Q$, $F$, and $V$ denote the main effects of calendar quarter, bar timeframe, and VAE variant; their pairwise and triple products denote interactions; $b_{A[i]} \sim \mathcal{N}(0, \sigma_A^2)$ is a random intercept for asset $i$; and $\varepsilon_i \sim \mathcal{N}(0, \sigma^2)$ is the residual. Factor $Q$ takes four levels (Q2 2025 through Q1 2026), factor $F$ four levels (5-minute, 15-minute, 30-minute, 60-minute), and factor $V$ three levels (VAE-I, VAE-II, VAE-III). All fixed-effect factors are encoded with sum-to-zero (effects) coding. Under this parameterisation the full model expands to

$$\begin{aligned} y_i \;=\; & \beta_0 + \sum_{q=1}^{3} \beta_q^Q\, Q_{iq} + \sum_{f=1}^{3} \beta_f^F\, F_{if} + \sum_{v=1}^{2} \beta_v^V\, V_{iv} + \sum_{q=1}^{3}\sum_{f=1}^{3} \beta_{qf}^{QF}\, Q_{iq}F_{if} \\ & + \sum_{q=1}^{3}\sum_{v=1}^{2} \beta_{qv}^{QV}\, Q_{iq}V_{iv} + \sum_{f=1}^{3}\sum_{v=1}^{2} \beta_{fv}^{FV}\, F_{if}V_{iv} + \sum_{q=1}^{3}\sum_{f=1}^{3}\sum_{v=1}^{2} \beta_{qfv}^{QFV}\, Q_{iq}F_{if}V_{iv} \\ & + b_{A[i]} + \varepsilon_i. \end{aligned} \tag{11}$$

Sum-to-zero coding yields three free contrasts for $Q$, three for $F$, and two for $V$, giving 3, 3, and 2 degrees of freedom for the respective main-effect terms. Statistical inference on fixed effects uses Type III $F$-tests with Satterthwaite (1946) degrees-of-freedom approximation. Overall model significance is assessed by a likelihood-ratio test (LRT) against a null model containing only the random intercept $b_{A[i]}$.

Because the five measures differ in their natural ranges, the regressand $y_i$ differs across models. Table 6 (Appendix A.5.) reports, for each measure, its raw range, the transformation applied, and the model type. Measures 1 and 2 are proportions bounded in $(0,1)$ and are transformed by the logit function; Measure 2 is first normalised by $H$ before logit is applied. Measure 3 is normalised by $W$ and requires no further transformation, as $\bar{w}/W$ does not approach either boundary in the data. Measure 4 concentrates near unity, so the Smithson–Verkuilen (2006) boundary correction is applied prior to logit to avoid numerical instability. Measure 5 may not exist — no sign change in $[1, H]$ — and is therefore handled by a two-part procedure: a binomial GLMM (Generalised Linear Mixed Model) for the existence indicator $Z = \mathbf{1}(h^* \text{ exists})$ (Part 1) and a linear mixed model for $\log(h^*/H)$ conditional on $Z = 1$ (Part 2).

## 4. Results

The analysis proceeds in two stages: establishing whether the anomaly signal carries any predictive content for cumulative returns, then characterising the shape of that predictive relationship across the full factorial design. The global null $H_0\colon \beta(w,h) \equiv 0$ is rejected for 100% of assets across all 48 experimental cells (4 timeframes × 4 quarters × 3 VAE methods). The estimated $\hat{\beta}(w,h)$ surface is statistically non-flat for all ten assets under all experimental conditions — this uniform outcome is the prerequisite for meaningful interpretation of the five scalar measures. The Ljung–Box test rejects serially uncorrelated residuals in 100% of cells. This is most plausibly attributable to persistent intraday microstructure dynamics — bid-ask bounce, order-flow imbalance, and volatility clustering — though model misspecification cannot be ruled out on the basis of the test alone. The unanimous rejection across all 480 asset-condition pairs, including the 5-minute bars where microstructure noise is most pronounced, is consistent with a structural rather than artifactual source.The five scalar measures are derived solely from the point estimates of $\hat{\beta}(w,h)$, not their standard errors; the mixed-model inference reported in Table 7 and Table 8 (Appendix A.5.) is unaffected by the first-stage autocorrelation structure.

The five scalar measures are analysed via the mixed model specified in equations (10)–(11), with regressand transformations documented in Table 6 (Appendix A.5.). Each model contains 48 structural fixed-effect parameters plus a random asset intercept. With $N = 480$ observations and ten random-effect levels, the residual degrees of freedom for the Satterthwaite $F$-tests are approximately 422. Models 1–4 and Model 5 Part 2 use Restricted Maximum Likelihood (REML) linear mixed models with Type III Satterthwaite $F$-tests; Model 5 Part 1 uses a binomial GLMM with Wald $\chi^2$ tests.

$H_1$ — persistence of cumulative return impact [$\delta(H/4)$ and $h_{0.50}$]. The grand-mean front-load ratio $\delta \approx 0.21$ (back-transformed from the logit-scale intercept of -1.318 via $\delta = 1/(1 + e^{1.318})$) lies below the proportional baseline of $1/4 = 0.25$ (intercept $p < .001$, Table 7). The cumulative return effect is therefore persistent — it continues to accumulate beyond the earliest horizons rather than dissipating at the anchor. The intercept of Model 2 is not significant ($p = .216$, Table 7), corresponding to $h_{0.50} \approx H/2$; the non-significance is itself the finding — the return effect reaches its half-life symmetrically in the centre of the horizon. A below-proportional early share and a centrally located half-life jointly confirm that the return effect builds gradually and persistently across the full response window. **$H_1$ is accepted.**

$H_2$ — distribution of predictive content [$\bar{w}/W$ and $U$]. The grand-mean $\bar{w}/W \approx 0.574$ ($p < .001$, Table 7) confirms that the centre of predictive mass sits in the latter half of the history window, giving more recent anomaly observations moderately greater weight. Back-transforming the Smithson–Verkuilen (2006) corrected logit gives $U \approx 0.94$ ($p < .001$, Table 7), close to unity; the predictor effect is near-uniformly distributed across $[0, W]$ — no single segment of the anomaly past dominates. The two measures jointly confirm that the entire history window is informative, with a statistically significant but modest tilt toward the most recent observations. **$H_2$ is accepted.**

$H_3$ — directional reversal within the forecast horizon [$h^*$]. Of the 480 experimental observations, 320 — 66.7% — exhibit a zero-crossing. The Part 2 LMM intercept ($p < .001$, Table 8) estimates $\log(h^*/H) \approx -1.014$, which back-transforms to $h^*/H \approx 0.36$. When a reversal occurs it does so in the first third of the horizon; the post-reversal lobe thereby spans roughly 64% of the horizon and dominates the absolute effect distribution, explaining both the below-proportional front-load ratio ($\delta < 1/4$) and the centrally located median horizon ($h_{0.50} \approx H/2$). **$H_3$ is accepted.**

Beyond the grand means, the factorial structure reveals systematic modulation. Quarter effects are significant for four of the five measures: $\delta$ ($p = .025$), $h_{0.50}$ ($p = .001$), $\bar{w}$ ($p = .024$), $h^*$ existence ($p = .008$), and $h^*$ timing ($p = .006$), but not for $U$ ($p = .481$). The near-uniform spread of predictive content across the history window is therefore invariant to market conditions; regime sensitivity is concentrated in the timing of the cumulative return profile, not in the distributional character of the predictor. The Q×Timeframe interaction is significant for $\delta$ ($p = .023$), $h_{0.50}$ ($p = .009$), and $\bar{w}$ ($p = .035$) but not for $U$ ($p = .425$), reinforcing that distributional uniformity is doubly stable — unaffected by regime and by any regime-frequency combination. Timeframe is the dominant driver for $\bar{w}$ ($p < .001$) and $U$ ($p < .001$);

the sampling frequency determines how predictive content is distributed across the trailing history, while the front-load ratio is insensitive to frequency ($p = .718$). The anomaly-scoring method is significant only for $U$ ($p = .040$) and $h^*$ timing ($p = .024$); the TF×Method interaction is additionally significant for $U$ ($p = .013$), indicating that the peer-context encoding's effect on distributional uniformity is modulated by bar frequency. The three core findings — $\mathrm{H}_1, \mathrm{H}_2, \mathrm{H}_3$ — are robust across VAE variants. Higher-order interactions are uniformly non-significant, confirming that the factorial structure is adequately described by main effects and twoway interactions. The asymmetry between quarter (governing timing) and timeframe (governing distribution) is itself a substantive finding: market regime and observation frequency govern qualitatively different dimensions of the anomaly-return relationship.

## 5. Limitations

The study draws on ten highly liquid S&P 500 constituents over a single twelve-month window concentrated in large-cap US equity and a structurally unusual period of recurring geopolitical stress; the documented anomaly-return patterns may not generalise to smaller or less liquid stocks, to non-US markets, or to periods of a qualitatively different character. The Amihud-based selection introduces a mild survivorship-type bias — stocks that became illiquid or were removed from the index during the sample period are excluded by construction — and all analyses are in-sample, so the predictive content of the anomaly signal cannot be verified in a genuine forecasting setting.

Several modelling choices are fixed on economic or practical grounds without formal sensitivity analysis. The Takens embedding parameters ($m = 3, \tau = 1$) are standard in the TDA literature but not theoretically optimal for the present data; alternative configurations could produce different topological graphs and thus different anomaly scores. The BallMapper radius $\hat{\varepsilon}$ is estimated from the MAD of the embedded point cloud — a data-adaptive but asymptotically ungrounded selector whose choice is consequential for which stocks are treated as peers at each snapshot. The predictor window $W$ and forecast horizon $H$ are fixed on economic grounds (Appendix A.3.) and held constant across all timeframes; the scalar measures are normalised by $W$ or $H$ and are interpretable in relative terms, but their absolute magnitudes depend on this choice. The three VAE variants use shallow architectures trained on a restricted hyperparameter grid; more expressive networks or finer tuning might yield anomaly scores with different distributional properties and thereby different estimated coefficient surfaces.

As reported in Section 4., the function-on-function regression residuals exhibit pervasive serial autocorrelation (100% of stocks across all 48 experimental conditions), which may overstate the precision of individual surface estimates and thus the unanimity of the zero-effect rejection. The mixed-model inference of Table 7 and Table 8 is not directly affected, since it operates on scalar summaries of the estimated surface rather than on the surface estimates themselves, but the first-stage significance results should be read with this caveat in mind. Additionally, the overall likelihood-ratio test for the front-load ratio model is non-significant ($p = .119$), indicating that the full fixed-effect structure provides only modest incremental explanatory power over asset-level heterogeneity alone for this measure; the individual Quarter and Q×Timeframe effects reported in Table 7 should accordingly be read as descriptive rather than confirmatory. Finally, all results are stated in gross log-returns; transaction costs, market impact, and bid-ask spreads are not accounted for. Quantifying net-of-cost economic significance presupposes that the signal has first been established as informationally non-trivial — which is what the present analysis demonstrates. Whether the anomaly-return link documented here survives in a net-of-cost, out-of-sample setting remains an open question for subsequent, operationally oriented research.

## 6. Conclusions

All three research hypotheses are accepted. Cross-sectional topological anomaly scores — defined as mismatches between a stock's observed intraday return pattern and what its peer-group topology leads one to expect — carry statistically significant and structurally coherent information about subsequent cumulative returns, confirmed across all ten assets, all four intraday frequencies, all four calendar quarters, and all three VAE peer-context variants. The return effect builds gradually from the anchor (first $H/4$ carrying only 21% of the total absolute impact), reverses at approximately one-third of the horizon ($h^*/H \approx 0.36$), and draws on the full anomaly history with near-uniform breadth ($U \approx 0.94$) and a modest recency lean ($\bar{w}/W \approx 0.57$). Together the five scalar measures describe a coherent temporal fingerprint: gradual accumulation, early sign change, and broadly distributed predictive content weighted toward recent history.

A key structural finding is the asymmetry between the two factorial drivers: calendar quarter — the proxy for market regime — governs the temporal dynamics of the return impact profile (when the effect accumulates and when it reverses), while bar timeframe governs the distributional character of the predictor effect (how evenly anomaly history contributes to prediction). The three VAE peer-context encoding strategies produce robustly consistent results — only fine-grained features of the coefficient surface are sensitive to the scoring method — confirming that the documented relationship is a structural property of the topological anomaly signal rather than a modelling artefact.

The anomaly score has three natural areas of practical application. It provides a real-time structural signal for market surveillance and risk monitoring: a stock with an elevated score is behaving in a way its peer group and the current market topology do not anticipate, directly relevant for position-level risk assessment. The typical reversal near $h^* \approx H/3$ suggests a natural exit horizon for contrarian intraday positions, though any such application must account for transaction costs and requires out-of-sample validation. Stocks whose anomaly profiles persistently differ from the peer-group pattern can be identified as structurally peripheral, which is informative for portfolio construction and concentration risk management.

Several extensions are left for future work. Out-of-sample evaluation is the most pressing priority. Extending the framework to a broader cross-section would test generalisability across the liquidity spectrum. Comparing the predictive content of the topological anomaly score with simpler cross-sectional measures — a return z-score or a factor-model residual — is an explicit direction for subsequent research, since establishing the signal's statistical existence is a necessary precondition for any such comparative evaluation. Multi-asset function-on-function regression and rolling-window VAE retraining to ensure adaptability to structural regime changes are further natural developments.

# Appendix

## A.0. Conceptual Framework

Figure 1

Three-stage research framework

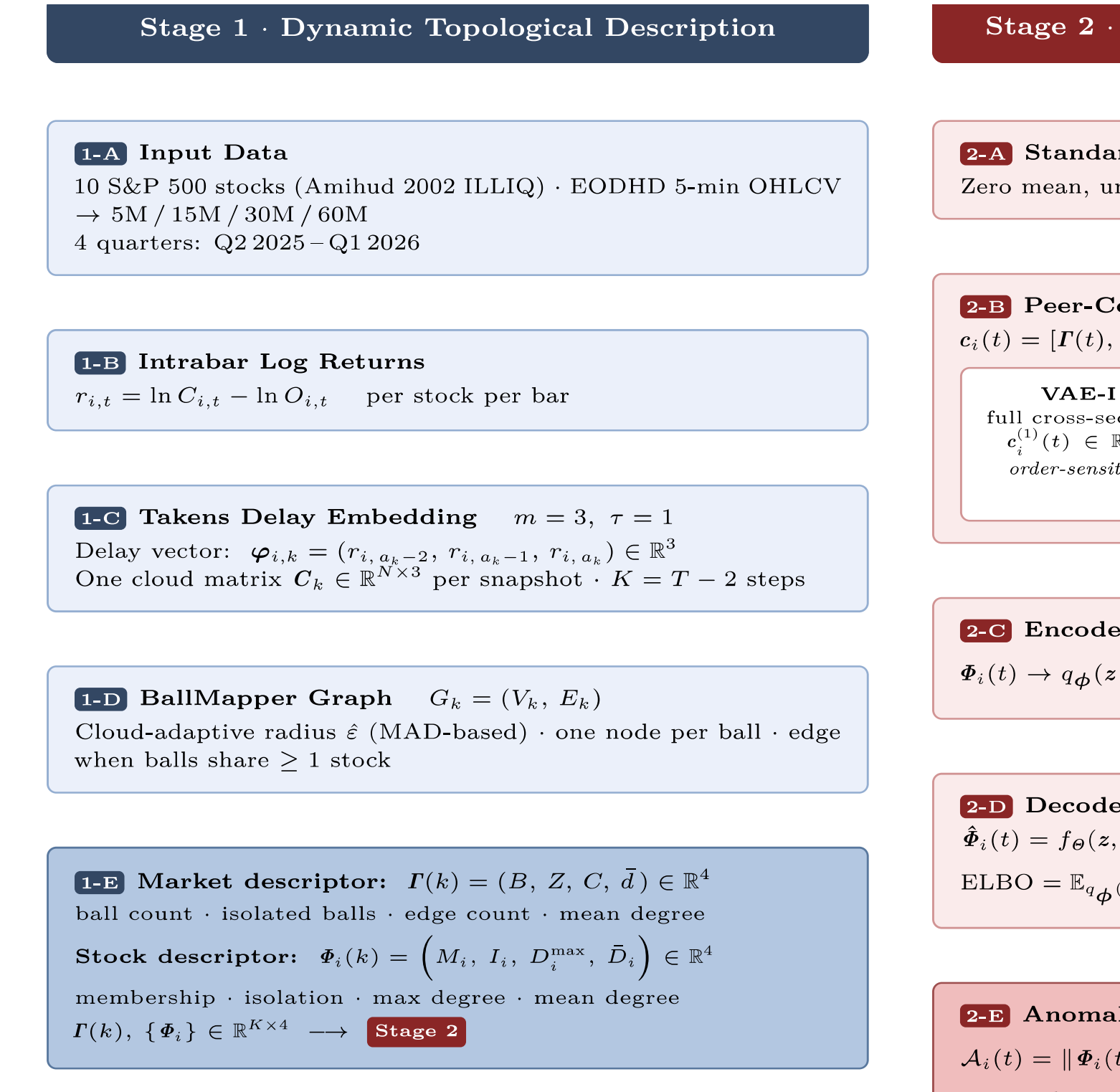

Stage 2 · Decoder-Conditional VAE (3 Variants)

2-A Standardisation
Zero mean, unit variance applied independently to $\Gamma$ and each $\Phi_i$

2-B Peer-Context Encoding
$c_i(t) = [\Gamma(t),$ peer summary$]$ — three conditioning variants:
VAE-I: full cross-section, $c_i^{(1)}(t) \in \mathbb{R}^{4N}$, *order-sensitive*
VAE-II: peer mean + SD, $c_i^{(2)}(t) \in \mathbb{R}^{12}$, *perm.-invariant*
VAE-III: attention-weighted, $c_i^{(3)}(t) \in \mathbb{R}^{8}$, *perm.-equivariant*

2-C Encoder (asset-specific, context-free)
$\Phi_i(t) \to q_\phi(z \mid \Phi_i(t)) = \mathcal{N}\left(\mu_\phi,\ \mathrm{diag}(\sigma_\phi^2)\right)$

2-D Decoder (context-conditional, $\beta$-VAE)
$\hat{\Phi}_i(t) = f_\theta(z,\ c_i)$
$\mathrm{ELBO} = \mathbb{E}_{q_\phi(z\mid\Phi_i(t))}[\log p_\theta(\Phi_i(t) \mid z,\ c_i)] - \beta\, D_{KL}\left(q_\phi \,\|\, \mathcal{N}(\mathbf{0}, \mathbf{I})\right)$

2-E Anomaly score per asset per snapshot:
$\mathcal{A}_i(t) = \|\Phi_i(t) - \hat{\Phi}_i(t)\|_2 + \beta \cdot D_{KL}\left(q_\phi \,\|\, \mathcal{N}(\mathbf{0}, \mathbf{I})\right)$
*structural surprise + latent irregularity*
Score matrix $A \in \mathbb{R}^{K\times N}$ ⟶ Stage 3

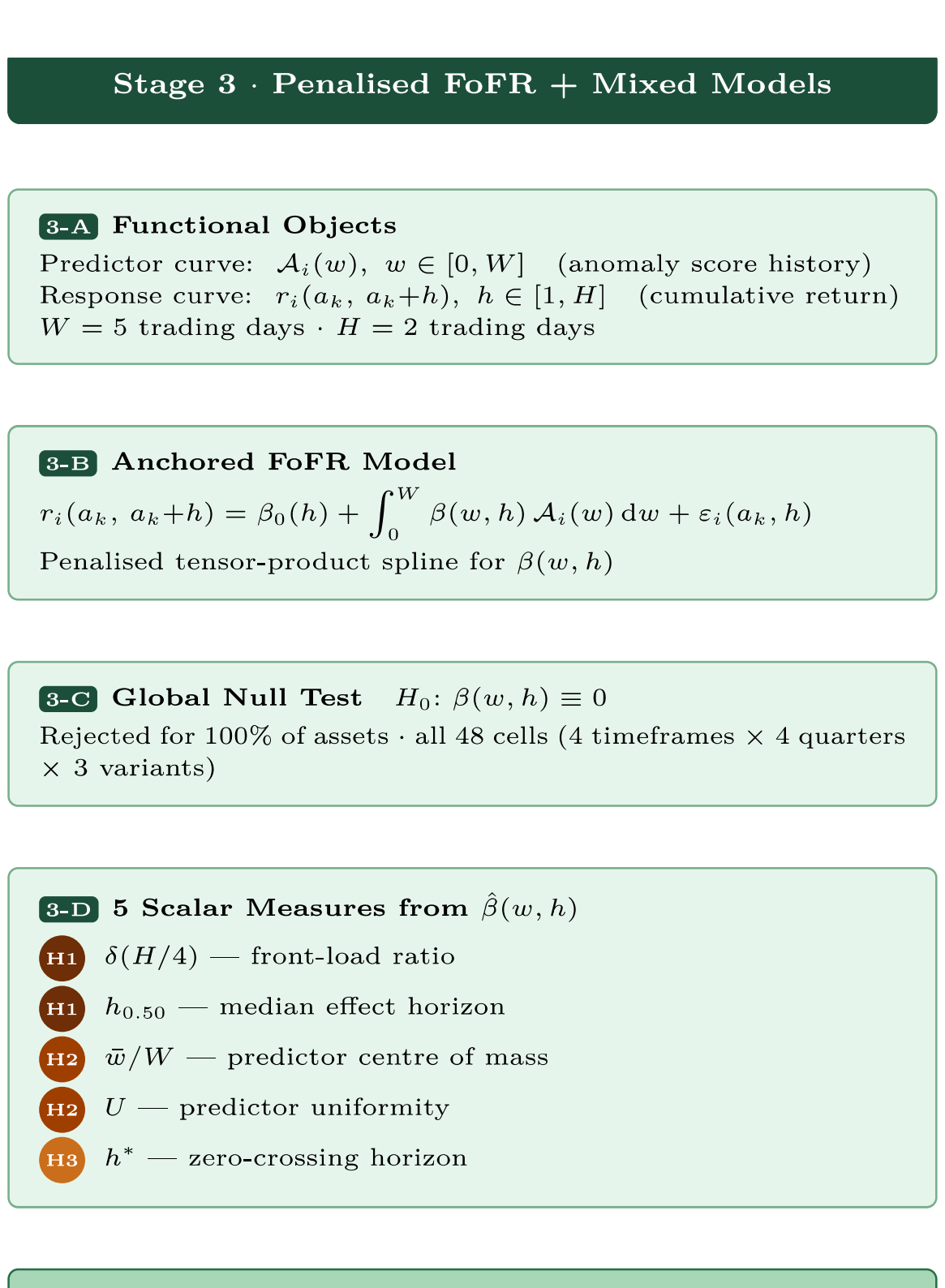

Note: Stage 1 (Dynamic Topological Description) is developed in Section 3.2.; Stage 2 (Decoder-Conditional VAE) in Section 3.3.; Stage 3 (Penalised FoFR + Mixed Models) in Section 3.4. Results across all three stages are presented in Section 4.

### A.1. Stock Selection and Liquidity Screening

The ten S&P 500 constituents used throughout the study are the ten stocks with the smallest time-series average Amihud (2002) illiquidity ratio over the sample period (2025-04-01 to 2026-03-31). For stock $i$ on day $t$ the daily Amihud ILLIQ is

$$ILLIQ_{i,t} = \frac{|r_{i,t}|}{P_{i,t} \cdot V_{i,t}}, \qquad r_{i,t} = \frac{Close_{i,t}}{Open_{i,t}} - 1,$$

where $P_{i,t}$ is the closing price and $V_{i,t}$ the share volume (dollar volume is expressed in millions). The per-stock measure is the mean over all valid trading days with at least 80% non-missing observations:

$$ILLIQ_i = \frac{1}{T_i} \sum_{t=1}^{T_i} ILLIQ_{i,t}.$$

Table 1
Top 10 Most Liquid S&P 500 Stocks — Amihud ILLIQ (2025-04-01 to 2026-03-31)

| Symbol | Company | $ILLIQ_i$ ($\times 10^{-6}$) | Avg Dollar Vol (bn) | Weight (%) |
|---|---|---|---|---|
| NVDA | NVIDIA CORP | 0.435 | 32.9 | 8.17 |
| TSLA | TESLA INC | 0.606 | 33.5 | 1.84 |
| AAPL | APPLE INC | 0.754 | 12.7 | 6.71 |
| MSFT | MICROSOFT CORP | 0.775 | 11.9 | 4.97 |
| AMZN | AMAZON.COM INC | 1.160 | 10.3 | 4.21 |
| META | META PLATFORMS INC CLASS A | 1.264 | 10.3 | 2.14 |
| GOOGL | ALPHABET INC CL A | 1.439 | 9.0 | 3.68 |
| GOOG | ALPHABET INC CL C | 2.188 | 5.9 | 2.94 |
| PLTR | PALANTIR TECHNOLOGIES INC A | 2.358 | 10.1 | 0.50 |
| AVGO | BROADCOM INC | 2.489 | 7.8 | 3.11 |

Note: Avg Dollar Vol (bn) = mean daily dollar volume in USD billions. Weight = S&P 500 index weight (%) at the time of data collection, included to characterise the market-capitalisation concentration of the sample (the ten stocks collectively account for approximately 38% of index weight, confirming large-cap dominance).

## A.2. Formal Definitions of Topological Descriptor Vectors

Table 2
Market-level 4D descriptor vector $\boldsymbol{\Gamma}(k)$

| Symbol | Definition | Interpretation |
|---|---|---|
| $B(k) = \lvert V_k \rvert \in \{1,2,\dots,N\}$ | Number of balls (graph nodes) | *Market fragmentation*. More balls = more asset dispersal in embedding space. High $B$ signals that the ten sampled S&P 500 constituents are spread across many distinct topological regions; low $B$ indicates concentrated co-movement. |
| $Z(k) = \lvert \{b \in V_k : \deg_k(b) = 0\} \rvert \in \{0,1,\dots,B(k)\}$ | Number of isolated balls (zero-degree nodes) | *Idiosyncratic cluster count*. Each isolated ball contains stocks whose trajectories share no geometric overlap with any other ball. High $Z$ at a snapshot signals stock-specific news or microstructural anomalies across multiple sub-groups. |
| $C(k) = \lvert E_k \rvert \in \left\{0,1,\dots,\frac{B(k)(B(k)-1)}{2}\right\}$ | Number of graph edges | *Inter-ball connectivity*. Rising $C$ indicates increasing topological coherence as overlapping balls link into larger connected components. High $C$ accompanies broadly correlated intraday return trajectories (e.g., macro shock episodes). |
| $\bar{d}(k) = \frac{1}{B(k)} \sum_{b \in V_k} \deg_k(b) \in [0, B(k) - 1]$ | Mean ball degree | *Average topological cohesion*. A single number summary of graph density, invariant to the absolute scale of $B(k)$. High $\bar{d}$ indicates a well-connected, hub-and-spoke market topology; low $\bar{d}$ with high $B$ indicates many weakly connected satellite balls. |

Table 3
Stock-level 4D descriptor vector $\boldsymbol{\Phi}_i(k)$

| Symbol | Definition | Interpretation |
|---|---|---|
| $M_i(k) = \lvert\mathcal{B}_i(k)\rvert \in \{1,2,\ldots,B(k)\}$ | Membership count: number of balls covering asset $i$ | *Topological boundary position*. $M_i = 1$ places asset $i$ inside a single, well-defined cluster; $M_i > 1$ means $i$ lies at the boundary between multiple overlapping regions, typical of hub stocks that co-move with several economically distinct sub-groups. |
| $I_i(k) = 1[\forall b \in \mathcal{B}_i(k): \deg_k(b) = 0] \in \{0,1\}$ | Isolation indicator (binary) | *Full idiosyncratic isolation*. $I_i = 1$ signals that every ball covering asset $i$ is disconnected from the rest of the graph at snapshot $k$-i.e., $i$ 's recent return trajectory is geometrically isolated from the entire cross-section. |
| $D_i^{\max}(k) = \max_{b \in B_i(k)} \deg_k(b) \in \{0,1,\ldots,B(k)-1\}$ | Maximum degree among all balls covering asset $i$ | *Access to the market core*. High $D_i^{\max}$ indicates that asset $i$ belongs to at least one highly connected hub ball, giving it structural proximity to the densest part of the graph. |
| $\overline{D}_i(k) = \frac{1}{M_i(k)} \sum_{b \in B_i(k)} deg_k(b) \in [0, B(k)-1]$ | Mean degree across all balls containing asset $i$ | *Average neighbourhood connectivity*. Complements $D_i^{\max}$ : a stock with high $\overline{D}_i$ is centrally embedded in dense regions throughout its topological neighbourhood, while high $D_i^{\max}$ with low $\overline{D}_i$ indicates peripheral adjacency to one hub only. |

Note: $\mathcal{B}_i(k) = \{b \in V_k : i \in \mathcal{A}_b\}$ where $\mathcal{A}_b$ is the set of stocks assigned to ball $b$. Takens embedding parameters: $m = 3, \tau = 1$. BallMapper radius $\varepsilon$ selected via MAD-based Gaussian approximation with coverage probability $\alpha = 0.68$; see Section 3.2.

### A.3. Predictor Window $W$ and Forecast Horizon $H$ : Derivation and Economic Rationale

**Principle: time-scale invariance**

$W$ and $H$ must represent the same economic time regardless of the bar granularity $\Delta t$ (5 min, 15 min, 30 min, or 60 min), so that results are directly comparable across timeframes.

**Trading-session lengthFormulas**

The S&P 500 continuous session runs from 09:30 to 16:00 ET: $T_{\text{session}} = 390$ min/day. The number of bars per trading day at granularity $\Delta t$ is therefore $b(\Delta t) = 390/\Delta t$.

**Formulas**

Two economic-time targets are fixed: $W = 5$ trading days (one calendar week) and $H = 2$ trading days. Converting to bars: $W = \lceil 5 \cdot b(\Delta t) \rceil$, $H = \lceil 2 \cdot b(\Delta t) \rceil$. The ratio $W/H = 5/2 = 2.5$ is approximately constant by construction (subject to ceiling rounding; see 60-min note in Table 4).

**Why these targets?**

- $W = 5$ days (1 week) captures the medium-frequency anomaly cycle: institutional rebalancing programmes typically operate over weekly intervals, index-related option expiry cycles create structural demand pressures on the same timescale, and earnings-window effects produce anomalous structural configurations in the BallMapper graph that persist for approximately one trading week before being absorbed. Setting $W$ shorter would fail to capture slow-building dislocations; setting $W$ substantially longer would include historical anomaly episodes whose structural influence on current returns is economically negligible.
- $H = 2$ days is long enough for the return impact of a structural anomaly to materialise through normal market mechanisms, but short enough to preclude compounding of unrelated macroeconomic price movements that would attenuate the detectable signal.
- The function-on-function regression maps the entire $W$-bar anomaly curve to the entire $H$-bar cumulative-return curve, so both endpoints must be set in consistent economic time; the bar-count formulas above guarantee this invariance across all four granularities.

Table 4
Bar counts for predictor window $W$ and forecast horizon $H$ across bar granularities

| **Time-frame $\Delta t$** | **$b(\Delta t)$ bars/day** | **$W = \lceil 5b \rceil$** | **$H = \lceil 2b \rceil$** | **$W/H$** | **Economic time $W$** | **Economic time $H$** |
|---|---|---|---|---|---|---|
| 5 min | 78 | 390 | 156 | 2.500 | 5 days | 2 days |
| 15 min | 26 | 130 | 52 | 2.500 | 5 days | 2 days |
| 30 min | 13 | 65 | 26 | 2.500 | 5 days | 2 days |
| 60 min | 6.5 | 33 | 13 | 2.538* | 5 days* | 2 days |

Note: *60-min case: $5 \times 6.5 = 32.5$, so $W = \lceil 32.5 \rceil = 33$; $2 \times 6.5 = 13$ exactly, so $H = 13$ with no rounding. The ratio $W/H = 33/13 \approx 2.538$ deviates marginally from 2.5 due to ceiling rounding of $W$ only.

## A.4. The Full Anchored Function-on-Function Model

The notation and roles of all quantities from formula (6) in Section 3.4. are summarized in Table 5.

Table 5
Notation for the anchored function-on-function regression model

| **Symbol** | **Domain** | **Role** |
|---|---|---|
| $k$ | $1, \ldots, K$ | Replication (anchor) index — one per snapshot |
| $h$ | $[1, H]$ | Horizon from anchor — response domain, identical for all replications |
| $w$ | $[0, W]$ | Integration variable — predictor domain; $w = 0$ oldest observation in window, $w = W$ anchor snapshot |
| $r_i(a_k, a_k + h)$ | Scalar for each $(k, h)$ | Cumulative return of asset $i$ from anchor $a_k$ to horizon $a_k + h$ |
| $A_i(w)$ | Scalar for each $w$ | Anomaly score of asset $i$ at past time $w$; the predictor curve for replication $k$ |
| $\beta_0(h)$ | Function of $h$ | Baseline post-anchor return curve (intercept function) |
| $\beta(w, h)$ | Bivariate surface on $[0, W] \times [1, H]$ | Coefficient surface: effect of anomaly at past time $w$ on cumulative return at horizon $h$ |
| $\varepsilon_i(a_k, h)$ | Scalar for each $(k, h)$ | Residual; independent across $k$ (subjects), correlated across $h$ within a replication |

### Reading the coefficient surface $\boldsymbol{\beta(w, h)}$

- Concentration near small $h$: anomaly history has a transient impact - the return effect decays quickly after the anchor.
- Spread across large $h$: impact is persistent - the anomaly predicts returns far into the future.
- Concentration near recent $w \approx W$: only the most recent part of the anomaly history matters for subsequent returns.
- Spread across $w$ : the full anomaly history (including distant past within the window) contributes to the return impact.

- Sign change across $h$: anomaly initially depresses returns but subsequently reverses - a mean-reversion structure directly visible from the surface shape.

### A.5. Regressand Transforms and Mixed-Effects Estimation Results

Table 6
Regressands in models (10)–(11): raw ranges, transformations applied, and model types.

| Measure | Raw range | Regressand $y_i$ | Model |
|---|---|---|---|
| $\delta(H/4)$ | (0,1) | logit($\delta(H/4)$) | LMM |
| $h_{0.50}$ | $[1, H]$ | logit($h_{0.50}/H$)$^\dagger$ | LMM |
| $\bar{w}$ | $[0, W]$ | $\bar{w}/W$ | LMM |
| $U$ | (0, 1), near 1 | logit($\bar{U}$); $\bar{U}$ = S − V correction of $U$ | LMM |
| $h^*$ | (1, $H$ ) or none | Part 1: $Z = \mathbf{1}$ ( $h^*$ exists) | Binomial GLMM |
| | | Part 2: log $(h^*/H)$ ∣ $Z = 1$ | LMM |

Note: LMM by REML; inference via Satterthwaite F-tests with sum-to-zero contrasts. Binomial GLMM fitted by maximum likelihood with logit link; inference via Wald $\chi^2$ tests. $S - V$ = Smithson-Verkuilen (2006) boundary shrinkage: $\bar{U} = (U(n-1) + 0.5)/n$ where $n$ is the total number of observations ( $n = 480$ in the full design). $^\dagger$Before applying logit, $h_{0.50}/H$ is clamped to $[10^{-4}, 1 - 10^{-4}]$ to prevent $\pm\infty$ at exact boundary values; no observed values fall outside this interval.

Table 7
Mixed-effects model p-values for Measures 1–4

| **Term** | $\boldsymbol{\delta(H/4)}$ | $\boldsymbol{h_{0.50}}$ | $\boldsymbol{\bar{w}}$ | $\boldsymbol{U}$ |
|---|---|---|---|---|
| Intercept | < .001 | 0.216 | < .001 | < .001 |
| Quarter | 0.025 | 0.001 | 0.024 | 0.481 |
| Timeframe | 0.718 | 0.094 | < .001 | < .001 |
| Method | 0.238 | 0.740 | 0.377 | 0.040 |
| Q × Timeframe | 0.023 | 0.009 | 0.035 | 0.425 |
| Q × Method | 0.939 | 0.501 | 0.583 | 0.104 |
| TF × Method | 0.084 | 0.227 | 0.210 | 0.013 |
| Q × TF × Method | 0.931 | 0.853 | 0.637 | 0.088 |
| Overall LRT | 0.119 | 0.009 | 0.003 | < .001 |

Table 8
Mixed-effects model p-values for Measure 5

| **Term** | **p (Exist.)** | **p (Timing)** |
|---|---|---|
| Intercept | 0.977 | < .001 |
| Quarter | 0.008 | 0.006 |
| Timeframe | 0.028 | 0.079 |
| Method | 0.797 | 0.024 |
| Q × Timeframe | 0.681 | 0.574 |
| Q × Method | 0.224 | 0.807 |
| TF × Method | 0.129 | 0.656 |
| Q × TF × Method | 0.944 | 0.590 |
| Overall LRT | 0.008 | 0.037 |

Notes for Table 7 and Table 8: Design: 3 methods (VAE-I, VAE-II, VAE-III) × 4 timeframes (5M, 15M, 30M, 60M) × 4 quarters (Q2-2025 through Q1-2026) × 10 assets = 480 observations. All models include a random asset intercept $b_{A[i]}$ (see formulas (10) and (11)) and sum-to-zero (effects) contrasts on all three fixed factors. Table 7: REML linear mixed models; p-values from Type III Satterthwaite F-tests. Responses (regressands): $\delta(H/4) -$ logit-transformed; $h_{0.50} -$ logit $(h_{0.50}/H)$, clamped to $[10^{-4}, 1 - 10^{-4}]$ before logit; $\bar{w} - \bar{w}/W$, no further transform; $U -$ logit of the Smithson-Verkuilen (2006) boundary-corrected value $\bar{U} = (U(n-1) + 0.5)/n$. Table 8, $p$ (Exist.): binomial GLMM fitted by maximum likelihood (Laplace approximation); p-values from Type III Wald $\chi^2$ tests. Table 8, $p$ (Timing): REML linear mixed model on the reversal subset ( $n = 320{,}66.7\%$ of 480 observations); p-values from Type III Satterthwaite $F$-tests; response log $(h^*/H)$. Overall LRT: likelihood-ratio test of the full fixed-effects model against a random-intercept-only null, both refitted by maximum likelihood. p-values are rounded to three decimal places; $< .001$ denotes $p < 0.001$.